\documentclass[a4paper,fleqn,usenatbib]{mnras}

\usepackage[T1]{fontenc}
\usepackage{ae,aecompl}

\usepackage{graphicx}	
\usepackage{amsmath}	
\usepackage{amssymb}	
\usepackage{bm}
\usepackage{lmodern}
\usepackage{pdflscape}
\usepackage{rotating}
\usepackage{adjustbox}
\let\oldhref\href
\renewcommand{\href}[2]{\oldhref{#1}{\hbox{#2}}}

\title[21-cm crosspower with 2dF galaxy survey at z $\sim$ 0.08]{Lack of 
clustering in low-redshift 21-cm intensity maps cross-correlated with 2dF galaxy densities}

\author[C. J. Anderson et al.]{C. J. Anderson,$^{1}$\thanks{E-mail: cjanderson23@wisc.edu}
N. J. Luciw,$^{2}$\thanks{E-mail: nluciw@cita.utoronto.ca}
Y.-C. Li,$^{3,4}$\thanks{E-mail: mailto:lixiating@gmail.com}
C.-Y. Kuo,$^{5,6}$
J. Yadav, $^{7,3}$
\newauthor
K. W. Masui,$^{8}$
T.-C. Chang,$^{9,6}$
X. Chen,$^{3,10}$,
N. Oppermann,$^{2,11}$
Y.-W. Liao,$^{6}$
\newauthor
U.-L. Pen,$^{2}$
D. C. Price,$^{12}$
L. Staveley-Smith,$^{13}$
E. R. Switzer,$^{14}$
P. T. Timbie,$^{1}$
\newauthor
and L. Wolz$^{15}$
\\
$^{1}$Department of Physics, University of Wisconsin Madison, 1150 University Ave, Madison WI 53703, USA\\
$^{2}$Canadian Institute for Theoretical Astrophysics, University of Toronto, 60 St. George St., Toronto Ontario, M5S 3H8, Canada\\
$^{3}$Key Laboratory of Computational Astrophysics, National Astronomical Observatory, Chinese Academy of Science, \\
\ 20A Datun Road, Beijing 100012, China\\
$^{4}$School of Chemistry and Physics, University of KwaZulu-Natal, Westville Campus, Private Bag X54001, Durban, 4000, South Africa\\
$^{5}$Department of Physics, National Sun Yat-Sen University, No. 70, Lienhai Rd., 804 Gushan Dist., Kaosiung City, Taiwan\\
$^{6}$Institute of Astronomy and Astrophysics, Academia Sinica, 11F of Astro-Math Building, AS/NTU, 1 Roosevelt Road Sec. 4,
Taipei 10617,\\
\ Taiwan\\
 $^{7}$Central University of Haryana, Jant-Pali, Mahendergarh - 123031, India\\
 $^{8}$Department of Physics and Astronomy, University of British Columbia, 6224 Agricultural Rd, Vancouver, BC, V6T 1Z1, Canada\\
 $^{9}$Jet Propulsion Laboratory, California Institute of Technology
4800 Oak Grove Dr, M/S 169-237, Pasadena CA 91109, USA\\
$^{10}$Center of High Energy Physics, Peking University, Beijing 100871, China\\
$^{11}$Dunlap Institute for Astronomy \& Astrophysics, 50 St.\ George St., Toronto Ontario, M5S 3H4, Canada\\
$^{12}$Department of Astronomy, University of California, Berkeley, 501 Campbell Hall \#3411, Berkeley CA 94720, USA\\
$^{13}$International Centre for Radio Astronomy Research, University of Western Australia, Crawley, WA 6009, Australia\\
$^{14}$NASA Goddard Space Flight Center, Greenbelt MD 20770, USA\\
$^{15}$School of Physics, University of Melbourne, Parkville, VIC 3010, Australia
}

\date{Accepted XXX. Received YYY; in original form ZZZ}

\pubyear{2017}

\begin{document}
\label{firstpage}
\pagerange{\pageref{firstpage}--\pageref{lastpage}}
\maketitle

\begin{abstract}
We report results from 21-cm intensity maps acquired from the Parkes radio telescope and cross-correlated with galaxy maps from the 2dF galaxy survey.  
The data span the redshift range $0.057<z<0.098$ and cover approximately 1,300 square degrees over two long fields.  
Cross correlation is detected at a significance of $5.18\, \sigma$.
The amplitude of the cross-power spectrum is low relative to the expected dark matter power spectrum, assuming a neutral hydrogen (HI) bias and mass density equal to measurements from the ALFALFA survey.  The decrement is pronounced and statistically significant at small scales.  At $k\sim1.5$ $ h \mathrm{\, Mpc^{-1}}$, the cross power spectrum is more than a factor of 6 lower than expected, with a significance of $14.8\,\sigma$.
This decrement indicates either a lack of clustering of neutral hydrogen (HI) 
, a small correlation coefficient between optical galaxies and HI
, or some combination of the two. 
Separating 2dF into red and blue galaxies, we find that red galaxies are much more weakly correlated with HI on $k\sim1.5$ $h \mathrm{\,Mpc^{-1}}$ scales, suggesting that HI is more associated with blue star-forming galaxies and tends to avoid red galaxies.

\end{abstract}




\section{Introduction}

Observations of neutral hydrogen (HI) provide a rich tool for understanding cosmology and astrophysics.  
Below $z \sim 6$, the majority of hydrogen is reionized by stellar radiation except for dense clumps of HI that are self-shielded from ionizing and Lyman radiation.  
These clumps, the densest of which are known as Damped Lyman Alpha systems (DLAs), are highly correlated with matter over-densities, where gravitational collapse provides the requisite hydrogen density for self-shielding.  
They are also thought to be crucial to star formation, since stars are unlikely to gravitationally collapse from warm ionized gas and are instead expected to evolve from cold neutral clouds, to molecular clouds, to collapsed stars \citep{prochaska2009non}.  
Measurements of neutral hydrogen therefore provide an opportunity to study star formation and to trace matter perturbations on large cosmic scales.

Several techniques exist to measure the HI density from redshifts of 0 to about 6.  
Above $z \sim 2.2$, the Lyman alpha line is redshifted to optical frequencies, and HI regions can be detected in absorption features from distant quasars \citep{prochaska2009non}.  
This technique has a maximum redshift of about 6, however, at which point the hydrogen neutral fraction seems to have been high enough for complete absorption of the quasar spectrum (Gunn-Peterson troughs). 
Below $z \sim 2.2$, the Lyman alpha line moves into the ultra-violet, and atmospheric scattering makes ground based measurements difficult.
A more suitable method at low redshifts is to use the 21-cm emission line of HI.

At very low redshifts, individual galaxies can be detected via blind searches for spikes in 21-cm emission.
The HI Parkes All Sky Survey (HIPASS) \citep{2004MNRAS.350.1195M} and Arecibo Legacy Fast ALFA (ALFALFA) \citep{2011AJ....142..170H} survey used this technique to detect individual galaxies out to $z \sim 0.04$ and $z \sim 0.06$ respectively.
However, blind galaxy searches become impractical at higher redshifts, since larger collecting areas are required to detect more distant galaxies.  
Higher redshift detections of individual galaxies, out to $z \sim 0.2$, can still be made by pointing telescopes at known galaxy positions, but this requires long integration times to overcome thermal noise.
The sky coverage of this method can be extended if one abandons the requirement of detecting individual galaxies.   
Instead, one can measure the 21-cm signal from the known locations and redshifts of many galaxies and boost the signal to noise by co-adding them.  
This technique is known as galaxy stacking, and it can yield high signal to noise measurements of the typical HI mass of optically selected galaxies. 
It is difficult to use galaxy stacking to determine the global HI content, especially at high redshift, since galaxy surveys will omit some of the optically dim HI galaxies.
However, stacking was used to infer the comoving HI density at $z < 0.13$ \citep{Delhaize:2013iea} and at $z=0.24$ \citep{lah2007h}.

A promising new technique is known as 21-cm intensity mapping.
Instead of cataloging individual galaxies, one can make low resolution three-dimensional maps of the large-scale structure (LSS) directly by detecting fluctuations in the aggregate 21-cm emission.  
Such surveys can be quickly carried out by radio telescopes and can in principle constrain the equation of state of dark energy by measuring the baryon acoustic oscillation feature to high accuracy \citep{PhysRevLett.100.091303}.
Because individual galaxies do not need to be identified, intensity mapping does not require telescopes with extremely large collecting areas in order to extend to high redshifts.
Another advantage of intensity mapping is that it is sensitive to emission from HI systems of all sizes.  
This feature contrasts with blind searches for HI galaxies, which are biased towards galaxies with large HI masses.
Cross-correlating intensity mapping surveys with galaxy surveys works similarly to galaxy stacking, but the aim is to measure clustering by analyzing the full correlation function or power spectrum of the 21-cm and galaxy maps.

The chief difficulty for 21-cm intensity mapping is the presence of radio point sources and free-free and synchrotron emission from the Galaxy.
These foregrounds are two to three orders of magnitude brighter than the HI signal at $z < 1$.
Fortunately, the inherent smoothness of the frequency spectrum of foregrounds \citep{Liu:2011ih} contrasts with the clumpy 21-cm signal that traces the redshift distribution of matter perturbations.
In the absence of instrumental effects, foregrounds could be removed by subtracting a slowly varying signal along the line of sight.  
However, telescopes convert the inherently smooth foregrounds to more complicated functions of frequency through imperfect bandpass calibration and frequency dependent beam patterns, and these instrumental effects necessitate the use of more sophisticated techniques to remove the foreground signal.  
In 2009, \cite{pen2009first} reported the first detection of cosmic structure using 21-cm maps from the HIPASS survey cross-correlated with the 6dF galaxy redshift survey \citep{jones20046df}.
The correlation function was measured over a small range of separations, from 0 to 3 Mpc.
In the northern hemisphere, the only 21-cm intensity mapping detection has come from HI maps from the Green Bank Telescope (GBT) cross-correlated with the WiggleZ and DEEP galaxy surveys at $z \sim 0.8$, using either Principal Component Analysis (PCA) \citep{chang2010intensity, 2041-8205-763-1-L20} or Independent Component Analysis (ICA) \citep{wolz2016erasing} to remove the foregrounds in the radio maps.
These measurements probed a much greater range of scales than the \cite{pen2009first} correlation function.

Here, we present the cross-power spectrum of 21-cm intensity maps from the Parkes Observatory correlated with the 2dF galaxy survey at $z \sim 0.08$.
This result is the first 21-cm intensity mapping result in the southern hemisphere to detect clustering at scales larger than 3 Mpc.
It demonstrates the robustness of the PCA foreground removal technique, which is shown to work for a multi-beam instrument with a different bandpass and beam shape from the GBT.
The redshift range of the Parkes measurement ($0.057<z<0.098$) is significantly lower than the GBT measurement ($0.6<z<1$), so a comparison of the two constrains the redshift evolution of the HI power spectrum.
The Parkes maps also probe smaller scales than the GBT measurement, which provides an opportunity to probe the mid-scale clustering characteristics of HI.

\section{Observations}

The Parkes 21-cm Multibeam Receiver \citep{1996PASA...13..243S} was used to map the two large contiguous fields of the 2dF Galaxy Redshift Survey \citep{Colless:1998yu} during a single week in late April and early May of 2014. 
During that week, 152 hours (1976 beam hours) of data on the two 2dF fields were collected.  
The Multibeam Correlator (MBCORR) backend was used with a 64 MHz bandwidth centered at 1315.5 MHz, a 62.5 kHz frequency bin size, and 2 second integration times.
Due to high variance at the edges of the band, the lowest 10 MHz and highest 4 MHz were removed from the final cross-power spectrum analysis described in section \ref{pow}.  
The results therefore cover a redshift range of $0.057<z<0.098$. 
High variance and odd bandpass shapes were immediately evident in data from two of the YY beams and one of the XX beams (XX and YY refer to the two orthogonal linear polarizations); data from these polarizations and beams were therefore not used in any of the analysis.
To minimize spurious signals from ground pickup, the telescope was positioned at a constant elevation angle during each field transit and scanned back and forth in azimuth as the field drifted through. 
Scans were made as the fields were rising and setting. 
The radio maps corresponding to the 2dF field near the North Galactic pole (NGP) cover roughly
4$^{\mathrm{h}}$30$'$ in right ascension and 11$^{\circ}$ in declination, centered at 12$^{\mathrm{h}}$ and 0$^{\circ}$.
The radio maps corresponding to the 2dF field near the South Galactic pole (SGP) cover roughly
6$^{\mathrm{h}}$30$'$ in right ascension and 7$^{\circ}$ in declination, centered at 0$^{\mathrm{h}}$40$'$ and -30$^{\circ}$.

\section{Data Analysis}
In this section we describe all the steps to go from raw data to calibrated maps in which spectrally smooth astrophysical foregrounds have been mostly removed, such that the remaining fluctuations are at the thermal noise level.
Subsection \ref{Mapmaking} describes RFI removal and mapmaking. Subsection \ref{bp} describes the bandpass and flux calibration procedure.  
Subsection \ref{foregrounds} describes the procedure for removing the smooth astrophysical foregrounds.
\subsection{Mapmaking}
\label{Mapmaking}
The raw data from MBCORR is stored in 3-minute blocks.
The first stage of our data analysis is a rough block by block cut to mitigate contamination by terrestrial sources of RFI. 
The high intrinsic spectral resolution of the data, 1024 channels across 64 MHz of bandwidth, allows for efficient identification and flagging of RFI. 
Individual frequency channels are flagged and removed if their variance, calculated across the duration of the block, is an extreme outlier compared to that of the other channels. 
Any RFI in a block that is not prominent enough to be flagged will contribute to a larger thermal noise estimation in the mapmaking stage, causing that block to be down-weighted when the map is made.

After RFI removal, the data are rebinned to 1 MHz bands (corresponding to a voxel depth of roughly 2.5 $h^{-1} \mathrm{\,Mpc}$ at band center).  
For each 3-minute block, the mean and slope in time are subtracted from the raw data, since these long time scale modes are contaminated by $1/f$ noise.
The timestream data are then converted to sky maps via an inverse-noise weighted chi-squared minimization.
This procedure has been used for CMB mapmaking (method 3 of \citealt{1538-4357-480-2-L87}), and it produces the maximum likelihood estimate of the sky map if the noise is Gaussian.

\begin{figure*}
\includegraphics[width=\textwidth]{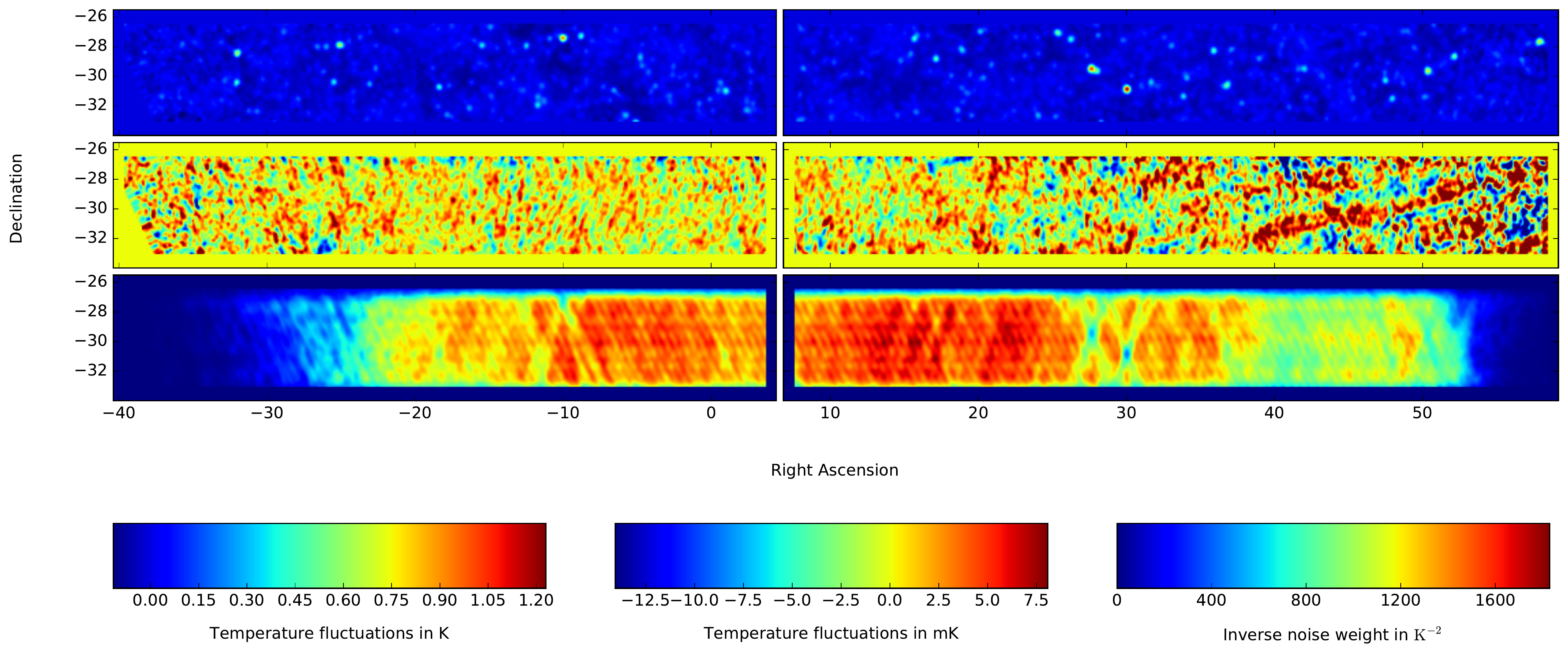}
\caption{The two 21-cm sub-maps that overlap the 2dF SGP field are shown at band center (the sub-maps that overlap the NGP field are not shown).  The three rows from top to bottom show:  the maps before any foreground modes are removed; the maps after 10 modes are removed; and the inverse noise weights, which are roughly proportional to the time spent observing each pixel.  
The color scales, from left to right, refer to the maps from top to bottom.  
All beams have been combined, and the resolution has been degraded to 1.4 times the original beamsize.  
The point sources and diffuse galactic foregrounds have been strongly suppressed in the foreground cleaned maps, and the scale of the remaining fluctuations is consistent with thermal noise. 
It should be noted that the fluctuations on the right ascension edges of the cleaned maps saturate the scale, but their magnitude is consistent with thermal noise and sparse coverage.
Some cross-hatched striping can be seen in the maps and weights due to the differing scan angles of the azimuthal scan strategy as the field rose and set.
The noise implied by the weights is higher than thermal noise because the mapmaker's noise estimation includes variance from the noise-cal measurement, residual RFI, and fluctuating foregrounds.}
\label{fig:2df1}
\end{figure*}

The noise covariance matrix is modeled in frequency-time space.
The model assumes no frequency correlations, and it allows for correlations in time, but only within each block.
Since there are no correlations between different blocks, the noise covariance matrix is estimated separately for each block.
Two pieces go into this estimate.
First, the thermal noise is estimated from the time variance of the data over the block and is placed on the diagonal.
Second, the mean and slope subtraction is accounted for in the noise model by adding large noise to orthogonal mean and slope modes in the time portion of the noise covariance.
This procedure allows the mapmaking algorithm to vary the mean and slope of each block to ensure maximum consistency between overlapping blocks.
Since data were collected on multiple days and at both rising and setting times, all the maps are built from a web of multiple overlapping blocks (the overlapping rising and setting scans can be seen in the inverse noise weights of Fig. \ref{fig:2df1}).
The multiple overlaps allow the mapmaking algorithm to distinguish real slowly-varying structure on the sky from spurious variations that are caused by $1/f$ noise or by the mean and slope subtraction.
The mapmaking algorithm also produces an inverse noise covariance matrix in map space, based on the noise model just described.  
The diagonal is kept for use as noise weights for the subsequent calibration, foreground removal, and power spectrum calculations.
For additional details on the mapmaking algorithm, see \cite{masui2013advancing}.

The mapmaking pipeline is based upon the pipeline used by \cite{2041-8205-763-1-L20}.
However, the larger size of the Parkes fields necessitated the development of parallel processing tools to overcome memory and speed issues.
Even with the parallelized code, it is necessary to break the two Parkes fields up into a total of four sub-maps.  
Map pixels are chosen to have a 0.08 degree width, which is approximately a third of the Parkes beam's full-width at half power (FWHP) of 0.25 degrees at band center.
This FWHP corresponds to approximately 1 $h^{-1} \mathrm{\,Mpc}$.
Each submap location is mapped separately for each beam and polarization.  
This is necessary because bandpass calibration is performed after map-making, and each beam and polarization has a slightly different bandpass shape.  
After bandpass calibration and foreground cleaning, the separate beam maps are co-added with inverse-noise weights.
Figure \ref{fig:2df1} shows the calibrated maps of the SGP field before and after foreground removal, along with the inverse-noise weights, after all beams have been co-added.

\subsection{Bandpass and Flux Calibration}
\label{bp}
A successful observation with a radiometer must (i) protect against fluctuations in the gain of the amplifiers, (ii) account for the bandpass spectrum (from frequency-dependent gain and bandpass filtering) that multiplies the true sky signal, and (iii) calibrate the total power by, for example, periodically observing astronomical sources of known brightness.  In this section, we describe how we implement these steps.

The MBCORR backend reduces gain fluctuations by injecting a flat broadband calibration signal, modulated by a 128 Hz square wave, into both polarizations of each beam.    
The signal is then amplified, downconverted using a local oscillator, filtered, digitized, and binned into 62.5~kHz frequency bins with a 2 second integration time.
At the same time, the full bandwidth is measured by total power detectors.  Synchronous demodulation of this signal at 128 Hz monitors the noise-cal power.
The digitally sampled data is then divided by this measured noise-cal power to provide gain stability.
Schematically, the data written by MBCORR is
\begin{equation}
D(\nu) = \frac{GB(\nu)P(\nu)P_{\mathrm{cal}\, 0}}{\langle GB(\nu)P_{\mathrm{cal}}\rangle_{64\, MHz}} = \frac{B(\nu)P(\nu)P_{\mathrm{cal}\, 0}}{P_{\mathrm{cal}}},
\end{equation}
where $G$ is the amplifier gain, $B(\nu)$ is the normalized bandpass shape, $P_{\mathrm{cal}}$ is the power of the noise-cal, $P_{\mathrm{cal}\, 0}$ is the pre-assumed power of the noise-cal, and $P(\nu)$ is the power measured by the digital sampler in the frequency band centered at $\nu$.
The extra power from the noise-cal adds about 1~K to the approximately 21~K system temperature.
Dividing by the small diode power introduces noise into the map which is highly correlated across all frequencies.  
Although this contribution to the noise is larger than the intrinsic thermal noise, it is removed in the first few modes of the foreground cleaning, described in \ref{foregrounds}. 

The bandpass has three effects on the data.  
The most significant effect is a systematic multiplication of the true sky spectrum by the average bandpass shape.
Secondly, fluctuations in the bandpass create noise in addition to the intrinsic thermal noise of the receiver.
Finally, there is an effect where the presence of bright point sources adds ripples to the bandpass shape due to standing waves between the receivers and the dish (see section 3.5 of \cite{Calabretta:2013yka} for a discussion of this effect for Parkes). 
Our bandpass calibration scheme aims only to estimate the average bandpass shape over the entire week long observation.
Though it may be possible to account for variation of the bandpass over time by computing multiple bandpasses for different times, this risks biasing the data. 

Each beam and polarization can have a different bandpass shape, in principle.
Our first step for estimating these bandpass spectra is to construct frequency covariance matrices between maps made with different beams. 
The covariance matrices are constructed as follows: let $\bm{M}_i$ represent a map produced by beam $i$, arranged as a matrix with frequency identified by the row and pixel by the column, and let $\bm{M}_j$ represent a map of the same region of the sky, from beam $j$.
Both maps are convolved to a common beam resolution, and a frequency covariance matrix $\bm{C}_{i,j}$ is constructed as
\begin{equation}
\bm{C}_{i,j} = \frac{\bar{\bm{W}}_i \circ \bm{M}_i (\bar{\bm{W}}_j \circ \bm{M}_j)^T}{\bar{\bm{W}}_i \bar{\bm{W}}_j^T},
\end{equation}
where $\bar{\bm{W}}_i$ represents a matrix of the inverse noise weights at each pixel for beam $i$, factorized into a separable function of angle and frequency. The factorization of the weights prevents frequency structure in the weights from influencing the covariance matrix. $\bar{\bm{W}}_i \circ \bm{M}_i$ is the element-by-element product of the map with the factorized weights.  
This covariance matrix is computed for all pairs of different beams, but only between common polarizations (thus avoiding thermal noise bias and spurious correlations due to polarized signal).  
We perform a singular value decomposition (SVD) on each covariance matrix.
\begin{equation}
\bm{C}_{i,j} = \bm{U}\bm{\Sigma}\bm{V}^T.
\end{equation}
The normalized left and right eigenvectors in $\bm{U}$ and $\bm{V}$ are ordered by their shared singular values, which are the entries of the diagonal matrix $\bm{\Sigma}$.
The singular values fall off rapidly, with the first singular value almost two orders of magnitude larger than the second.
The projections of the first left and right eigenvectors onto their maps reveal that the first modes are due to diffuse emission over the full region of the maps.
Therefore, the first eigenvector of $\bm{V}$ represents an estimate of the bandpass shape of beam $j$ multiplied by the frequency spectrum of the strong diffuse emission in the map, which will be dominated by galactic synchrotron radiation.
Similarly, the first eigenvector of $\bm{U}$ represents the diffuse emission spectrum multiplied by the bandpass of beam $i$.
This interpretation is also motivated by SVD analysis of GBT maps.
For the GBT maps, which are bandpass calibrated with a noise diode before mapmaking, the first eigenvector has been found to capture the frequency spectrum of the diffuse synchrotron emission across the map \citep{0004-637X-815-1-51}.
Let $\hat{u_i}(\nu)$ represent the average first eigenvector of beam $i$, computed by averaging the first left eigenvector of $\bm{C}_{i,j}$ over each other beam, $j\neq i$.  
We assume a Galactic synchrotron spectral index of -2.7, which must be divided out to attain an estimate of the bandpass from $\hat{u_i}(\nu)$. We compute a bandpass estimate for beam $i$ thusly:
\begin{equation}\label{eq:bp}
\hat{B}_i(\nu) = \frac{\hat{u}(\nu)\nu^{2.7}}{\langle \hat{u}(\nu)\nu^{2.7} \rangle_{64\, MHz}}.
\end{equation}
Equation \ref{eq:bp} is the bandpass estimate that is used for each beam and polarization and which is computed separately for each of the four sub-maps.  The bandpass estimates are similar but clearly distinct for each beam, as one would expect since the amplifiers and filters are similar but not identical.  The variance in each beam's bandpass estimate from Equation \ref{eq:bp} is negligible in any one region.  The standard deviation across the four sub-maps is slightly less than $3\%$.

Next, calibration of the total flux is achieved via periodic on/off scans of the astronomical point source J1939-6342 and beam mapping scans made with the point source PKS0043-42.  
In the on/off scans, the power of the switching noise-cal, $\hat{P}_{\mathrm{cal}}$, is measured in Janskys by comparison to J1939-6342. 
The consistency of these measurements verifies the stability of the noise-cal to $\sim1\%$ over the full week of observations. 
Measurements from \cite{1965AuJPh..18..329P} and \cite{1969AuJPh..22...79G} found the flux of J1939-6342 to be $16.4 \pm 1.0$ Jy, which translates to an absolute calibration uncertainty of $6.1\%$.
To convert from Janskys to Kelvin, we require the forward gains of the beams. 
These are obtained by mapping the beams with PKS0043-42 and fitting the shapes to a Gaussian.  Utilizing these fits and the fact that the total effective collecting area for any antenna is
\begin{equation}
\int_{4\pi}A_e(\theta,\phi)d\Omega = \lambda^2 ,
\end{equation}
we find that each beam has a forward gain in our band of $G = 0.79\, \mathrm{K \, Jy^{-1}}$ with an uncertainty of $5\%$.  Combining the point source flux, noise-cal stability, and forward gain errors in quadrature, we estimate a total flux uncertainty of $8\%$.

With all these factors in hand, we apply a bandpass and flux calibration to the maps as follows:
\begin{equation}
\bm{M}_i^{C}(\theta, \nu) = \frac{G\bm{M}_i(\theta, \nu)\hat{P}_{\mathrm{cal}}}{P_{\mathrm{cal}\, 0} \hat{B}_i(\nu)}.
\end{equation}
The inverse noise weights produced by the mapmaker are also calibrated, following a similar procedure:
\begin{equation}
\bm{W}_i^{C}(\theta, \nu) = \frac{\bm{W}_i(\theta, \nu)P_{\mathrm{cal}\, 0}^2 \hat{B}_i(\nu)^2}{G^2\hat{P}_{\mathrm{cal}}^2}.\end{equation}

In preparation for the next stage of the analysis, we average maps from the two linear polarizations to form unpolarized maps.  The 13 beams are then averaged, using their inverse noise weights, into four groups.  A: beams 1-3, B: beams 4-6, C: beams 7-9, and D: beams 10-13.  Although group D appears to have data from more beams, two of the YY and one of the XX beams from that group are not included, because of high variance and strange bandpass shapes.

\subsection{Foreground removal}
\label{foregrounds}
Extragalactic point sources and the Milky Way produce synchrotron emission that is two to three orders of magnitude brighter than the 21-cm signal.
In the absence of instrumental effects, these foregrounds are thought to be spectrally smooth and easily separable from the signal \citep{Liu:2011ih}, occupying just a few spectral degrees of freedom.  
In practice, bandpass instability, frequency dependent beam response, and leakage of polarized foregrounds into unpolarized signal all conspire to impart a complicated frequency structure onto the originally smooth foregrounds. 
More importantly, these effects can mix the local angular structure of the map into frequency structure at each pixel. 
Therefore, we assume that the spectral structure of the foregrounds
is not known a priori and must be determined from the data.
Since the foregrounds are the dominant component in the maps, we determine these foreground modes via a Principal Component Analysis of the maps.

We construct frequency-frequency covariance matrices as
\begin{equation}
\bm{C}_{\mathrm{A},\text{B}} = \frac{\bar{\bm{W}}_\text{A} \circ \bm{M}_\text{A} (\bar{\bm{W}}_\text{B} \circ \bm{M}_\text{B})^T}{\bar{\bm{W}}_\text{A} \bar{\bm{W}}_\text{B}^T}
\end{equation}
using pairs of calibrated maps and factorized calibrated weights.
Before a covariance is constructed, all frequency slices of the maps are convolved to a common beam resolution based on a frequency-dependent Gaussian beam model.
To the extent that we can accurately model the beam, this step prevents the frequency dependent beam size from altering the shape of the SVD modes and from coupling angular variation in the maps into frequency variation.
For each of our four sub-maps, we form six covariance matrices using all pairs of different beam groups: AB, AC, AD, BC, BD, CD.
We then perform a singular value decomposition on each covariance matrix
\begin{equation}
\bm{C}_{\text{A},\text{B}} = \bm{U}\bm{\Sigma}\bm{V}^T
\end{equation}
where the columns of $\bm{U}$ and $\bm{V}$ are the orthonormal spectral modes of $\bm{M}_\text{A}$ and $\bm{M}_\text{B}$ respectively, ordered by their shared singular values.
The singular values in $\bm{\Sigma}$ roughly correspond to the squared power present in the map from each frequency mode.  
Because foregrounds dominate the power in the map and are much more coherent between different frequencies than the 21-cm signal, the modes with the largest singular values (the principal components) are considered to be foreground modes.
By only correlating maps made by different groups of beams, we have avoided any thermal noise bias in the determination of these foreground modes.
Our foreground cleaning procedure projects out a fixed number of these spectral foreground modes from all pixels of the 21-cm maps.  
The number of foreground modes to remove is determined via simulations, as detailed in Section \ref{pow}.
Our SVD foreground removal process is formally described in
\cite{0004-637X-815-1-51} and is used in the analysis of \cite{2041-8205-763-1-L20}.
Note that those examples use or assume a single receiver, which necessitates correlating separate season maps to avoid the noise bias that we avoid here by correlating maps made from different beams.

Although the frequency modes with the highest singular values are dominated by foregrounds, there will inevitably be loss of 21-cm signal when these modes are removed from the maps.
This loss of power must be accounted for in the cross-power spectrum by the application of a transfer function, as described in Section \ref{pow}.  

\section{Power spectrum estimation} \label{pow}
In this section, we describe our method for estimating the cross-power spectrum between our foreground cleaned 21-cm intensity maps and the 2dF galaxy overdensity maps.  
In order to compensate for signal loss from the foreground cleaning and to estimate error bars, we simulate the dark matter power spectrum and draw mock galaxy and HI maps from this simulation.  
We then use a Monte Carlo method to calculate the error bars, running 100 of these simulated galaxy and HI maps through our power spectrum pipeline.
Subsection \ref{sims} describes these simulations, and Subsection \ref{power} describes the estimation of the cross-power spectrum and its errors.
\subsection{Simulations} \label{sims}
  
The non-linear dark matter power spectrum at $z \, = \, 0.08$ is computed using the HALOFIT routine (based on the method described in \cite{smith2003stable}) of the Cosmic Linear Anisotropy Solving System (CLASS, \cite{2011JCAP...07..034B}). 
We use Planck 2015 \citep{ade2016planck} values for the input cosmological parameters.
In comoving Cartesian k-space, 100 random Gaussian dark matter realizations are drawn from this dark matter power spectrum for each of the four sub-map regions.
In order to introduce redshift space distortions (RSDs), the z-axis is chosen to be the line-of-sight direction (flat-sky approximation).
The Fourier amplitudes are then multiplied by the angle-dependent factor $b(1 + \beta \mu^2)/\sqrt{1+(k\mu\sigma_v/H_0)^2}$, where $b$ is the bias of the matter tracer, $\mu$ is the cosine of the angle between the line-of-sight and the $k$-vector, $\sigma_v$ is the dispersion of the velocity field, $H_0=100$  $h \mathrm{\,km\,s^{-1}\,Mpc^{-1}}$, and $\beta = f(z)/b$, with $f(z)$ being the dimensionless growth rate.  
Non-linear RSDs, known as fingers-of-god, are introduced on small scales through the denominator and linear RSDs, known as the Kaiser effect, are implemented through the numerator \citep{1994MNRAS.267.1020P, 1987MNRAS.227....1K}.
We create an HI fluctuation map and an optical galaxy overdensity map from each of the 100 dark matter realizations.
For the dispersion of the velocity field, we use $\sigma_v=500$ $\textrm{km/s}$, from the fit of \cite{hawkins20032df} to the 2dF correlation function. 
We choose $b_\text{HI} \sim 0.85$, supported by \cite{0004-637X-718-2-972} at the redshifts of interest.  
The optical galaxy bias $b_\text{g}$ is expected to be $\sim 1$ for 2dF galaxies at the redshifts of interest \citep{Cole:2005sx}, so a bias of 1.0 is used for the galaxy over-density maps.
Each HI map is also multiplied by an assumed average HI brightness temperature, $T_b=0.064 \text{ mK}$.  This value comes from the ALFALFA survey \citep{2010ApJ...723.1359M} measurement of the comoving HI density, $\mathrm{\Omega_{HI}}$, through the equation \citep{chang2010intensity}
\begin{align} \label{eqn:temp}
T_\text{b}(z) = 0.39 \: \frac{\Omega_{\text{HI}}}{10^{-3}} \left[ \frac{\Omega_{\text{m}} + \Omega_{\Lambda}(1+z)^{-3}}{0.29} \right]^{-1/2} \left[ \frac{1+z}{2.5} \right]^{1/2} \; \text{mK}.
\end{align}
All cosmological quantities are in units of today's critical density. 
$\mathrm{\Omega_{\text{m}}}$ and $\mathrm{\Omega_{\Lambda}}$ are the the matter and dark energy densities at the present epoch.
We use Planck 2015 \citep{ade2016planck} values again for these.
The k-space maps are then converted to regular comoving coordinates.
These steps imply a cross-power spectrum in redshift space given by Equation \ref{eqn:power}, with a cross-correlation coefficient of unity.

For the next stage of the analysis, the simulated maps must be converted to telescope coordinates of frequency, right ascension, and declination.
This requires a fiducial cosmology (we again use Planck 2015) and a gridding scheme. 
In order to preserve the z-axis as the line-of-sight direction, the conversion of transverse lengths into angular distances in right ascension and declination assumes a constant radial distance, independent of the redshift -- for this purpose, the radial distance that halves the volume of the survey is chosen.  
The maps are then interpolated onto evenly spaced frequency intervals.
An unclustered mock galaxy catalog, following the survey selection function, is added to each galaxy over-density map to approximate the effect of galaxy shot noise.
A set of galaxy density maps without this shot noise contribution is also kept.
One set of the HI fluctuation maps is kept unaltered, and a second set is convolved with a Gaussian beam of width 1.4 times the largest Parkes beam, equal to the resolution of the common-beam-convolved real radio maps.

 \subsection{Cross-power spectrum} \label{power}

The procedure for estimating the cross-power spectrum of a pair of HI fluctuation and galaxy overdensity maps is as follows.
First, the maps are multiplied by their weights: the selection function for the galaxy map, and the inverse noise weights for the HI map.
Then, they must be converted from right ascension, declination, and frequency coordinates to physical comoving coordinates.
This conversion is the reverse of the procedure described in the second paragraph of Subsection \ref{sims}.  The z-direction is chosen to be the line of sight.  To convert angular distances to transverse distances, the same radius is used for all redshifts -- the radius that halves the volume of the survey.
This approximation results in a slightly distorted map that occupies a cube in Cartesian comoving coordinates, with the z-direction corresponding to the line-of-sight.
Lastly, the map is interpolated onto evenly spaced coordinates in the z-direction.

The 3D power spectrum can then be estimated as
\begin{equation}
\bm{P} = \frac{\bm{\mathcal{F}}\left(\bm{M}_{\text{HI}} \circ \bm{C}^{-1} \right) \circ \bm{\mathcal{F}}\left(\bm{M}_{\text{g}} \circ \bm{S} \right)^*}{N}\, ,
\end{equation}
where $\bm{\mathcal{F}}(\bm{X})$ represents the 3D Fourier transform of map $\bm{X}$ in comoving coordinates, $\circ$ denotes element-wise multiplication, $\bm{M}_{\text{HI}}$ is the HI fluctuation map, $\bm{M}_\text{g}$ is the galaxy number overdensity map, $\bm{C}$ is the diagonal of the noise covariance of the observed 21-cm map, $\bm{S}$ is the galaxy survey selection function, and $N = \sum_{i,j,k} \left(\bm{{C}}^{-1} \circ \bm{S} \right)_{ijk}$ is a normalizing factor, where $i,j,k$ denote the map voxel coordinates.  The 3D power spectrum is then averaged over the azimuthal angle with respect to the line-of-sight and binned to form a 2D power spectrum in $(k_{\bot},k_{\parallel})$ space.  The 2D cross-power spectrum of the real HI and galaxy maps is  a robust estimate of the cross-power spectrum, because thermal noise and residual foregrounds are not expected to correlate with the optical survey and tend to average to zero due to the azimuthal average and binning.  However, the cross-power will be systematically low due to signal loss from foreground cleaning and convolution with the telescope beam.  This loss of power must be accounted for by the transfer function, which is calculated from the simulated maps.  

To estimate the transfer function, we follow the procedure described in detail by \cite{0004-637X-815-1-51}.  
The transfer function is calculated in $(k_{\bot},k_{\parallel})$ space, which is a natural basis given that the foreground cleaning operates in the line-of-sight $k_{\parallel}$ direction and the beam convolution operates in the $k_{\bot}$ direction.
Schematically, the transfer function is given by
\small
\begin{align}
\begin{split}
&\bm{T}_{\alpha} = 
\biggl< \bm{Q} \bigg( \bm{\mathcal{F}} \left\{ \left[ \bm{\Pi}_{\text{M+s}}  ( \bm{M}_{\text{pks}} + \bm{X}_{\text{HI}}^\text{c} ) - \bm{\Pi}_{\text{M}}  \bm{M}_{\text{pks}}\right] \circ \bm{C}^{-1} \right\} \circ\\
& \bm{\mathcal{F}} \left\{\bm{X}_{\text{g}}^\text{sn} \circ  \bm{S}\right\}^*  \bigg)_{\alpha} \biggr> \quad \bigg/ \quad \biggl<  \bm{Q} \big( \bm{\mathcal{F}}\{\bm{X}_{\text{HI}} \circ \bm{C}^{-1}\} \circ \bm{\mathcal{F}}\{{\bm{X}_{\text{g}} \circ  \bm{S}}\}^* \big)_{\alpha} \biggr> ,
\end{split}
\end{align}
\normalsize
where the operator $\bm{Q}()_\alpha$ represents the azimuthal averaging of the 3D k-vectors into $(k_{\bot},k_{\parallel})$ bins indexed by $\alpha$, $\bm{M}_{\text{pks}}$ is the real Parkes 21-cm map, $\bm{X}_{\text{HI}}$ is the simulated HI 21-cm fluctuation map, $\bm{X}_{\text{HI}}^\text{c}$ is that same simulated map convolved with a Gaussian beam, $\bm{X}_{\text{g}}$ is the simulated galaxy over-density map, $\bm{X}_{\text{g}}^\text{sn}$ is the simulated galaxy over-density map with galaxy shot noise, and the angled brackets indicate averaging over the 100 simulations.
The $\bm{\Pi}$ operator represents the removal of the spectral foreground modes, which are calculated via the SVD procedure described in Section \ref{foregrounds}.
The $\bm{\Pi}_{\text{M}}$ operator removes foreground modes that are determined from the common beam convolved real map, and the $\bm{\Pi}_{\text{M+s}}$ removes foreground modes that are determined from the sum of the convolved simulated HI map and the common-beam-convolved real map.  
The numerator of the transfer function can be considered our best estimate of the cross-power we would expect to measure from the Parkes instrument if our model for the simulated power spectrum were correct.  
It includes the effects of signal loss from foreground cleaning, residual foregrounds, thermal noise, galaxy shot noise, and beam convolution. 
When cleaning the simulated map, it is essential to calculate the foreground modes from the sum of the real map and the simulated signal, since the HI signal perturbs the measured foreground modes \citep{0004-637X-815-1-51}.
The subtraction of $\bm{\Pi}_{\text{M}} \; \bm{M}_{\text{pks}}$ decreases the variance in the estimation of the numerator.
The denominator of the transfer function can be considered our best estimate of the simulated cross-power that would be measured in the absence of foregrounds, thermal noise, and shot noise. Therefore, the transfer function represents the fraction of signal retained in the cross-power after foreground removal and beam convolution.

Our estimate of the observed 2D cross-power spectrum, now compensated for signal loss by the transfer function, is
\begin{equation}
\widehat{\bm{P}_{\alpha}} = \frac{\bm{Q} \bigg( \bm{\mathcal{F}} \left\{ \left[ \bm{\Pi}_{\text{M}}  \bm{M}_{\text{pks}}\right] \circ \bm{C}^{-1} \right\} \circ\ \bm{\mathcal{F}} \left\{\bm{M}_{\text{g}}  \circ  \bm{S}\right\}^*  \bigg)_{\alpha} }{N \bm{T}_{\alpha}}.
\end{equation}
Similarly, we can calculate each simulated recovered 2D cross-power spectrum:
\small
\begin{equation}\label{eq:sim_power}
\bm{P}_{\alpha}^{\text{sim}} = \frac{\bm{Q} \bigg( \bm{\mathcal{F}} \left\{ \left[ \bm{\Pi}_{\text{M+s}} ( \bm{M}_{\text{pks}} + \bm{X}_{\text{HI}}^c \right) ] \circ \bm{C}^{-1} \right\} \circ\ \bm{\mathcal{F}} \left\{\bm{X}_{\text{g}}  \circ \bm{S}\right\}^*  \bigg)_{\alpha} }{N \bm{T}_{\alpha}}.
\end{equation}
\normalsize
The mean of this quantity over 100 simulations represents the 2D cross-power spectrum we would expect to estimate, after compensating for signal loss with the transfer function, if our model for the cross-power were correct.
The variance provides an estimation of the expected errors, incorporating thermal noise, foreground residuals, sample variance, signal loss from foreground cleaning, and galaxy shot noise.
The number of foreground modes to remove is chosen to minimize this variance.  
If too few modes are removed, residual foregrounds boost the variance.
If too many modes are removed, most of the signal is also removed and the small transfer function in the denominator boosts the errors.
We find that removing 10 SVD modes minimizes this variance.

To display our final results, we average the power spectrum to 1D bins.
The average to 1D is weighted by the inverse variance of each $k$-bin across the 100 simulated recovered 2D power spectra.
The observed cross-power spectrum, cleaned by removing 10 SVD modes, is shown in Fig. \ref{fig:crosspow}; we display only 1D Fourier modes for which we have full 2D angular coverage.
The uncertainty assigned to each bin of the final observed 1D power is the corresponding standard deviation calculated from the 100 simulated 1D power spectra. 
The full covariance of the binned 1D power spectrum over the 100 simulations is also checked; we find no significant correlations, so the error bars at each point of figure \ref{fig:crosspow} are independent.
We list the standard deviations of the 2D power spectra in Appendix \ref{appB}.  The inverse square of these are the weights used to bin to 1D power spectra. 
Each of the four sub-maps is analyzed independently of the others --  the final result is an average of these.

As a null test for correlations between residual foregrounds and 2dF galaxies, we randomly shuffle the redshift slices of our 21-cm maps and compute the cross-power spectrum between these shuffled maps and the 2dF maps -- we find the cross power is consistent with zero on all scales of interest.

\section{Results and Discussion}

\begin{figure*}
\includegraphics[width=\textwidth]{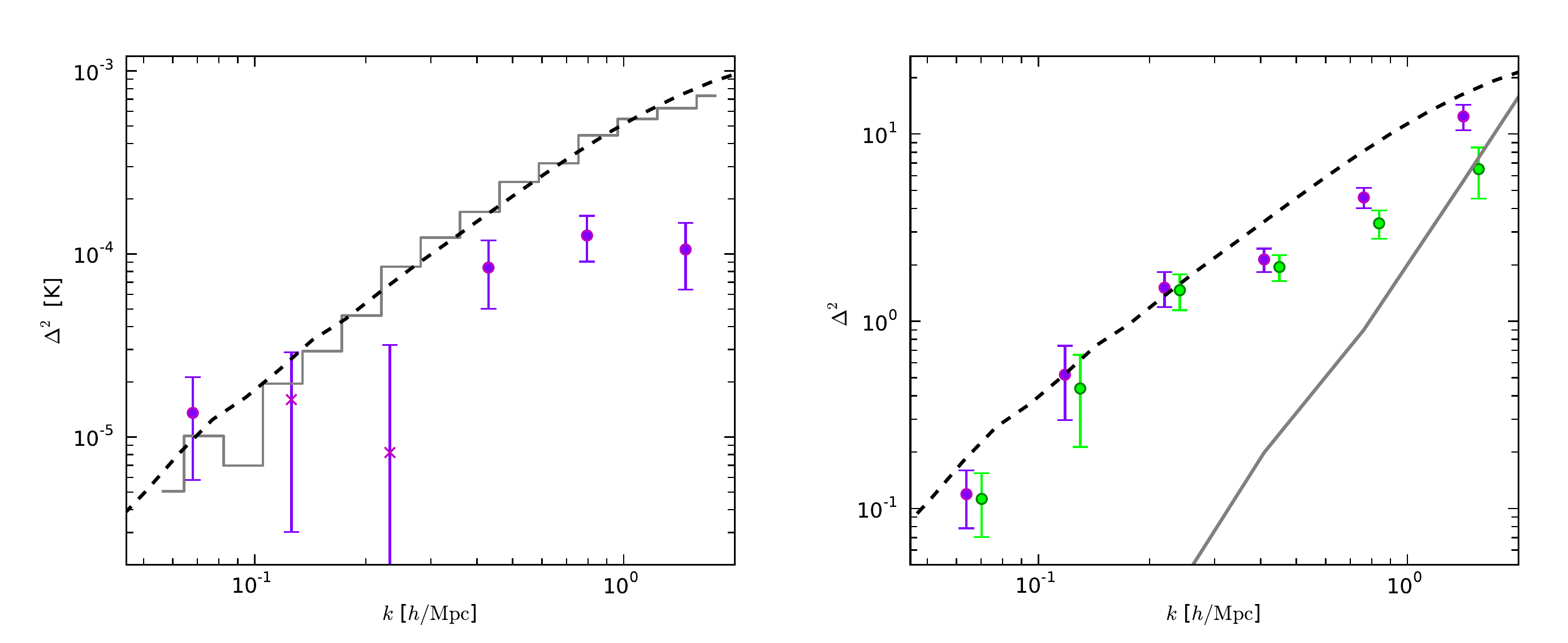}
\caption{Left: Observed 1D cross-power averaged over the four Parkes fields and cleaned by removing 10 SVD modes. A circle denotes positive power and a $\times$ denotes negative power. The grey line is the mean of the simulations, for which we assume $b_\text{HI} = 0.85$ and $T_\text{b} = 0.064 \text{ mK}$, given by the ALFALFA measurement of $\Omega_\text{HI}$ \citep{2010ApJ...723.1359M}. The dashed black line is the corresponding dark matter power spectrum scaled as Eq \ref{eqn:power}. Plotted error bars are 1-$\sigma$, derived from the Monte Carlo simulations described in section \ref{pow}.
Right: The purple points are the average of the auto-power spectra of the 2dF galaxies in the regions that overlap our Parkes maps. Errors are the standard deviation of the mean over the four sub-map regions.  The dashed black line is the simulated dark matter power spectrum.  The solid grey line is the expected shot noise signal, simulated from 100 unclustered mock catalogs that follow the survey selection function.  The green points are the 2dF auto-power data minus the simulated shot noise.}
\label{fig:crosspow}
\end{figure*}
The observed galaxy-HI cross-power spectrum is shown on the left-hand side of Fig. \ref{fig:crosspow}.  
A simple model for the expected cross-power spectrum is the simulated CLASS HALOFIT dark matter power spectrum, with an assumed scale-independent galaxy bias $b_{\text{g}}=1.0$ \citep{Cole:2005sx}, HI bias $b_{\text{HI}}=0.85$ \citep{0004-637X-718-2-972}, and mean 21-cm brightness temperature $T_\text{b} = 0.064 \text{ mK}$ \citep{0004-637X-718-2-972}.  
The dashed black curve displays this model.  In true comoving space, this model cross-power is 
\begin{equation}\label{eq:simple_ps}
P_{\text{HI},\text{g}}(k) = T_{{b}}\, b_{\text{HI}}\, b_{\text{g}}\, r\, P_{\delta \delta}(k),
\end{equation}
where $P_{\delta \delta}(k)$ is the CLASS dark matter power spectrum, and $r$ is the galaxy-HI correlation coefficient, defined as 
\begin{equation}
r = \frac{P_{\text{HI},\text{g}}}{\sqrt{P_{\text{HI},\text{HI}}P_{\text{g},\text{g}}}}.
\end{equation}
The true cross-power spectrum will also contain shot noise, but we do not attempt to model it, since it depends on the typical HI content of 2dF galaxies, which is unknown. 
It should be noted that the galaxy-HI correlation coefficient obeys the Schwarz inequality: $-1\leq r \leq 1$.  
A value of $\lvert r \rvert < 1$ would mean that HI fluctuations and galaxy over-densities are not simple multiples of each other in Fourier space; one would find phase differences or fluctuating amplitude differences within each k-bin of their Fourier transformed maps.
For the dashed black curve of the model, we assume $r=1$ on all scales.  

However, using the Halo Occupation Distribution (HOD) model \citep{berlind2002halo,zheng2005theoretical}, one can deduce that the expected behavior of the cross-correlation coefficient is scale-dependent.
In the HOD model, the power spectrum is the sum of three terms:
\begin{equation}
P(k) = P^{2\mathrm{h}}(k) + P^{1\mathrm{h}}(k) + P^{\mathrm{SN}}(k).
\end{equation}
$P^{2\mathrm{h}}(k)$ is the 2-halo term, which comes from matter tracers that occupy separate halos; $P^{1\mathrm{h}}(k)$ is the 1-halo term, due to clustering of matter tracers within the same halo; and $P^\mathrm{SN}(k)$ is the shot noise term.
On large scales, the 2-halo term is dominant. 
On intermediate scales, the 2-halo term falls off, and the 1-halo term and shot noise begin to dominate the power spectrum.
On the smallest scales, the 1-halo term will also fall off due to the finite extent of the halos, and shot noise will dominate.

Let us now consider the degree to which HI and galaxies will correlate for each of these terms.
Since galaxies and HI should both be contained within halos, and the distribution of halos will trace the underlying dark matter density field, we expect $r^{2\mathrm{h}}\approx1$.
On the other hand, it is likely that HI and optically selected galaxies have a tendency to occupy different halos.
\cite{hess2013evolution} studied the group membership of over 740 overlapping optical galaxies from SDSS and HI galaxies from ALFALFA.
They found that only 25\% of HI galaxies appear to be associated with an optically identified group, compared to half of optical galaxies.
This tendency for HI to occupy different halos suggests that both shot noise and 1-halo clustering may not correlate between HI and optically selected galaxy populations: we expect $r^{1h}<1$ and $r^{SN}<1$.
Therefore, $r$ is thought to be close to unity on large scales and to fall off on smaller scales, as shot noise and 1-halo clustering begin to dominate the power spectrum. 

Now, let us analyze our measured cross-power spectrum.
As previously indicated, the dashed black line in the left panel of Fig. \ref{fig:crosspow} shows the power spectrum of Equation \ref{eq:simple_ps} (which includes no shot noise term) with $r=1$.
Redshift space distortions, which modify the power spectrum according to Equation \ref{eqn:power}, are also included; this curve is a binning of this distorted 2D cross-power to 1D with isotropic weights.
The grey line shows the signal we'd realistically expect to measure if our model were accurate.
It represents the average recovered cross-power spectrum from the 100 simulated galaxy and HI map pairs, including the effects of our window function, thermal noise, residual foregrounds, galaxy shot noise, compensation for signal loss, and anisotropic weighting (see Appendix \ref{appB}).
The error bars on the data points show the standard deviation of these 100 simulations.
The deviations of the data from the grey line in the left panel of Fig. \ref{fig:crosspow} indicate disagreement with the simple model of Equations \ref{eq:simple_ps} and \ref{eqn:power}.
In summary, the cross-power is well below the expectation of our model at all scales except for the largest scale ($k\sim 0.07~ h\mathrm{\,Mpc^{-1}}$).
The two negative points between $0.1~ h \mathrm{\,Mpc^{-1}}$ and $0.3~ h \mathrm{\,Mpc^{-1}}$ are slightly troubling, since there is no physical reason to expect an anti-correlation between HI and optical galaxies on large scales.  However, the deviations from zero are not very significant, and those points are consistent with low positive clustering, or even the simulation line at $\sim 2 \sigma$.
However, the decrement from the model is statistically significant at the two highest $k$ points.  
At $k\sim 0.8~ h\mathrm{\,Mpc^{-1}}$, the signal is approximately 34\% of the model, and the significance of the decrement is $6.83\,\sigma$. 
At $k\sim 1.5~ h\mathrm{\,Mpc^{-1}}$, the signal is approximately 15\% of the model, and the significance of the decrement is $14.8\,\sigma$.  We estimate the total statistical significance of the detected cross-power by calculating $\chi^2$, using the error bars from our 100 simulations.  This rules out the null hypothesis of zero cross-power to an equivalent Gaussian significance of about $3.75\,\sigma$.  Since most of our data is on small scales where the linear bias model may fail, we do not attempt to fit any curve to our data.

For comparison, the right panel of Fig. \ref{fig:crosspow} shows the auto-power spectrum of the 2dF galaxies that overlap our Parkes fields.
We estimate the shot noise contribution to this power spectrum by averaging the power spectrum of 100 unclustered mock catalogs that follow the survey selection function; the shot noise estimate is the grey curve.
The purple points show our calculated 2dF auto-power spectrum, and the green points show this same power spectrum after subtracting the estimated shot noise contribution.
The galaxy power spectrum after shot noise removal shows a similar decrement in clustering at high $k$ to the HI-galaxy cross-power spectrum, but the effect is not as drastic at the two highest $k$-bins.
Our 2dF power spectrum roughly agrees with the graphed or tabulated 2dF power spectra of \cite{Cole:2005sx} and \cite{tegmark2002power} at the points where they overlap, but the overlap is mostly at low $k$.  The smallest scales analyzed in those papers are $k\sim 0.185~\mathrm{h\,Mpc^{-1}}$ and $k\sim 0.6~\mathrm{h\,Mpc^{-1}}$ respectively.  
The analysis of \cite{percival20012df} displays the ratio of the 2dF power spectrum to model fits, extending to $k\sim 1~\mathrm{h\,Mpc^{-1}}$.  Their plots show a decrease in 2dF power relative to the models at small scales, which is similar to the effect that we find.  However, they ascribe this effect to aliasing from coarse binning.  
Due to the rather coarse frequency binning of our maps, it is conceivable that aliasing is also responsible for some of the low power we observe at small scales.
In order to test this, we bin the galaxy maps first to the same frequency resolution as our HI maps and then with a factor of eight finer frequency resolution and calculate the power spectrum for both cases.  
We find a nearly identical galaxy power spectrum, indicating that aliasing is not an issue.

\begin{figure*}
	\includegraphics[width=\textwidth]
{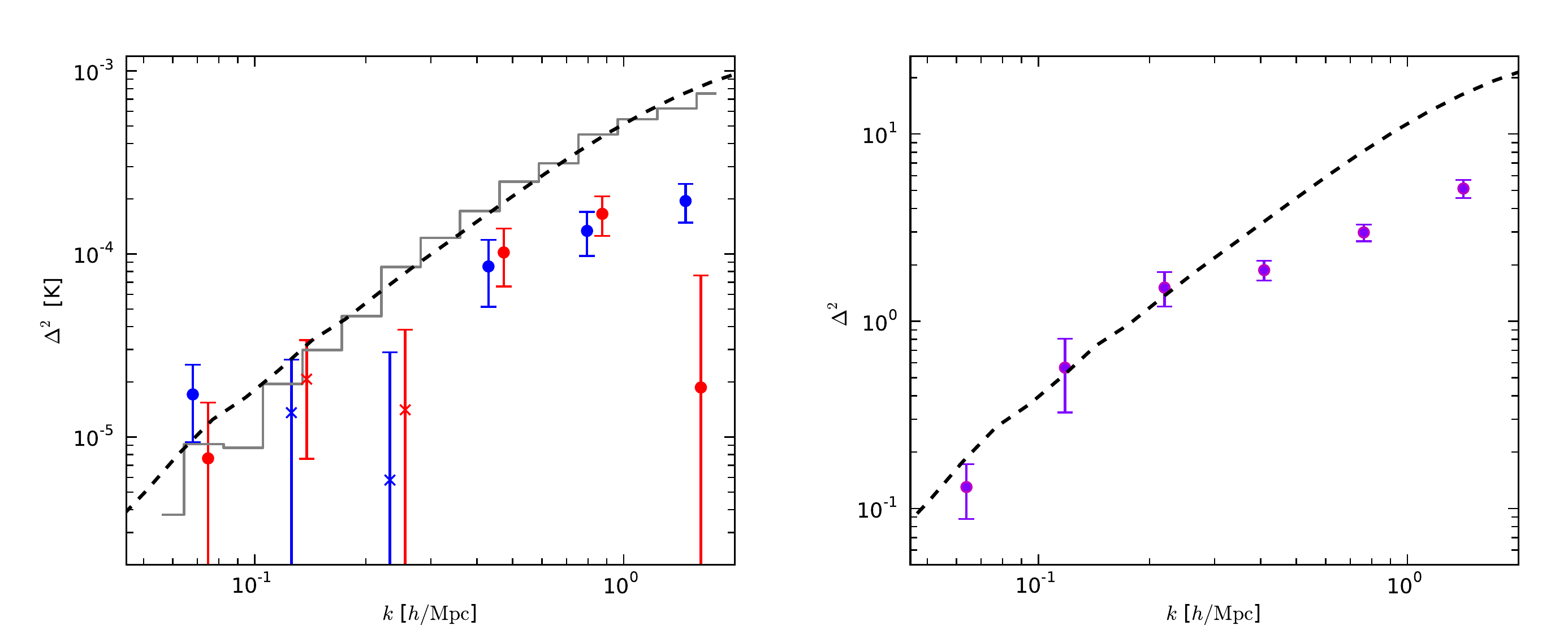}
	\caption{Similar to figure \ref{fig:crosspow}, but the 2dF galaxies have been split into red and blue populations.  The 100 simulations, used to calculate the error bars and the transfer function, have their galaxy counts adjusted to the appropriate number for the red and blue splits, to capture the effect of increased shot noise.  Left: Observed 1D cross-power spectrum between the HI maps and the red and blue 2dF galaxies, averaged over the four Parkes fields and cleaned by removing 10 SVD modes. The red cross-power points are slightly offset on the k-axis, for ease of reading.  Right: Observed 1D 2dF red-blue galaxy cross-power spectrum averaged over the regions that overlap our Parkes maps.  Errors are the standard deviation of the mean over the four sub-map regions.}
    \label{fig:redblue}
\end{figure*}

It is likely that some of the low HI-galaxy clustering that we see on small scales is due to the correlation coefficient dipping below 1, since shot noise and 1-halo clustering become more prominent in the power spectrum at high $k$.
A reasonable way to test this is to split the galaxies by color, under the hypothesis that the HI content of red galaxies is lower than that of blue galaxies.
If this hypothesis is true, we would expect the HI-red galaxy correlation coefficient to drop more rapidly at small scales than the HI-blue galaxy correlation coefficient. 
Following the k-corrected color splitting method used in \cite{Cole:2005sx}, we split the 2dF galaxies into red and blue populations and analyze the HI-galaxy cross-power spectra. In our fields, red galaxies account for approximately a third of the total 2dF population, and blue galaxies account for two thirds.
The cross-power spectra with the blue and red galaxies are shown in the left panel of Figure \ref{fig:redblue}.
The two cross-powers are quite similar, except that there is significantly more power at $k\sim 1.5~ h\mathrm{\,Mpc^{-1}}$ in the HI cross-power with the blue galaxies compared to the HI cross-power with the red galaxies.  
The statistical significance of the difference is about $2.4\,\sigma$.
This result favors the picture that the cross-power spectrum of HI and optical galaxies at $k\sim 1.5~ h \mathrm{\,Mpc^{-1}}$ is dominated either by shot noise or the 1-halo term.
Shot noise and 1-halo clustering is seen more strongly when correlating HI with blue galaxies because blue galaxies contain a much larger fraction of the HI and are more likely to occupy the same halos.  
A $\chi^2$ test of the cross-power between the HI signal and the blue 2dF galaxies rules out the null hypothesis of zero cross-correlation to an equivalent Gaussian significance of $5.18\,\sigma$.

The results of our HI cross-power spectra with red and blue galaxies are qualitatively consistent with the findings of \cite{papastergis2013clustering}.  Their analysis of the projected cross-correlation function and auto-correlation function of red and blue SDSS galaxies and HI-selected ALFALFA galaxies reveals that the HI-blue cross-correlation coefficient is close to unity at all scales.   
On the other hand, the HI-red cross-correlation coefficient is unity at large separations, but it begins to drop at separations smaller than $\sim 5~ h \mathrm{^{-1}\,Mpc}$, indicating that the presence of a red galaxy decreases the probability of finding a nearby HI-galaxy, relative to statistically independent dark matter tracers.
As noted by \cite{papastergis2013clustering}, this result may reflect the fact that red galaxies tend to preferentially inhabit high density halos \citep{zehavi2011galaxy}, which usually have lower fractions of HI gas, as seen in studies of individual groups and clusters \citep{haynes1986pattern,solanes2002three,hess2013evolution}, hydrodynamic simulations \citep{villaescusa2016neutral}, and empirical fits to the halo mass function \citep{padmanabhan2016constraints}.  
A lack of HI mass for many red galaxies is also found in simulations by \cite{wolz2016intensity}, using a semi-analytical galaxy formation model run on the Millennium simulation.  A cross-power spectrum analysis of this simulation reveals a similar scale-dependence and color-dependence to the correlation coefficient.  

The analysis of \cite{papastergis2013clustering} suggests that optically selected blue galaxies and HI galaxies tend to occur in the same environments or perhaps are the same galaxies.  To test this, we plot the cross-power spectrum of red and blue galaxies on the right panel of Fig. \ref{fig:redblue}, suspecting that it may be similar to the cross-power spectrum of HI and red galaxies.  In fact we find that the cross-power spectrum of red and blue galaxies is more similar to the full galaxy auto-power spectrum with shot noise subtracted.  This makes sense, since the blue and red galaxies are disjoint sets.  The side-by-side comparison of red-blue galaxy cross-power spectrum and the HI-red cross-power spectrum in Fig. \ref{fig:redblue} suggests that the overlap of HI with red galaxies is even weaker than the overlap of red and blue galaxies.  

Three points can be drawn from our results.  First, the small-scale clustering amplitude is much lower than the HALOFIT prediction, as seen in both the galaxy-HI power spectrum and the shot-noise subtracted galaxy auto-power spectrum.  Second, the galaxy-HI cross-correlation coefficient is scale-dependent and color-dependent, probably due to 1-halo clustering and shot noise.  HI appears to be much more strongly associated with blue galaxies than red galaxies.  Third, HI-galaxy clustering may also be somewhat suppressed at $0.1~ h \mathrm{\,Mpc^{-1}}$ and $0.3~ h \mathrm{\,Mpc^{-1}}$ scales, though the statistical significance of this is not high.  

\section{Conclusions}
 
We measure the cross-power spectrum between foreground cleaned 21-cm intensity maps and 2dF galaxy maps at $z\sim0.08$.  The cross-power spectrum is lower than the expected signal, if one assumes the ALFALFA values for the HI density and HI bias.  The decrement compared to the model has high statistical significance at $k\sim 1 ~ h\text{Mpc}^{-1}$.  Despite the smaller than expected signal, the null hypothesis of no correlated clustering between HI and galaxies can be ruled out to a significance of $5.18\, \sigma$, using the cross-power spectrum of the 21-cm maps and the blue 2dF galaxies.
The detection demonstrates the effectiveness of the SVD method for 21-cm foreground removal.
A splitting of the galaxies by color reveals that an extreme observed decrement in power at $k\sim 1.5 ~ h \text{Mpc}^{-1}$ is likely due to a scale-dependent correlation coefficient between HI and red galaxies that falls off sharply as shot noise and the 1-halo term begin to dominate the clustering power.  This supports the picture that the HI content of blue galaxies is much greater than the HI content of red galaxies.   
Constraints from the 21-cm auto-power spectrum will be the subject of future work.

\section*{Acknowledgements}

The authors would like to thank Ettore Carretti for his assistance in operating and understanding the Parkes telescope.
Computations were performed on the GPC supercomputer at the SciNet HPC Consortium.
CJA acknowledges support from the Wisconsin Space Grant Consortium.
CJA and PTT acknowledge support from NSF award AST-1211781.  XC acknowledges support from MoST 863 grant 2012AA121701,
NSFC Key Project 11633004, CAS project QYZDJ-SSW-SLH017.




\bibliographystyle{mnras}
\bibliography{refs} 




\appendix

\section{Redshift-space cross-power spectrum} \label{appA}

In redshift space, the power is distorted \citep{2015MNRAS.447..234W} and can be modeled as
\begin{align} \label{eqn:power}
\begin{split}
P_{\text{HI},\text{opt}}(k_{\parallel}, k_{\perp},z,\mu) = &T_{\text{b}}\; b_{\text{HI}}\; b_{\text{opt}}\; r \quad \times \\ &\frac{(1+\beta_{\text{HI}}\, \mu^{2})(1+\beta_{\text{opt}}\, \mu^{2})}{1 + (k\mu  \sigma_v/H_0)^2}P_{\delta \delta}(k_{\parallel}, k_{\perp}) \; ,
\end{split}
\end{align}
where $\beta_i = f(z) \, / \, b_i$, $f(z)$ is the dimensionless growth rate, $\mu$ is the cosine of the angle between the line-of-sight and the $k$-vector, and $\sigma_v$ is the dispersion of the velocity field. The numerator of the fraction arises from linear theory of infall into over-densities \citep{1987MNRAS.227....1K} and the denominator from non-linear theory relating the small-scale velocity field to a field characterized by dispersion $\sigma_v$ \citep{1994MNRAS.267.1020P}. Expanding the above in terms of the Legendre polynomials yields the power monopole, given by the expansion coefficients corresponding to the zeroth-order Legendre polynomial.
\newpage
\section{2D Weights}\label{appB}
The following two tables are the standard deviations of the 2D $k$-bins across 100 simulations per field, propagated through an average of the cross power over the four Parkes fields; the corresponding inverse variance is used as weights for the 2D to 1D average. The first column in each table gives $k_\parallel$ and the first row in each table gives $k_\perp$. \\

\label{weightapp}
\begin{adjustbox}{angle=90}
\centering
\resizebox{0.90\paperwidth}{0.075\paperheight}{
\setlength\tabcolsep{4pt}
\begin{tabular}{lcccccccccccccc}\hline
\multicolumn{1}{c}{$_{k_\parallel \backslash ^k{_\perp}}$} & \textit{3.03e-02} & \textit{3.72e-02} & \textit{5.63e-02} & \textit{6.93e-02} & \textit{8.53e-02} & \textit{1.05e-01} & \textit{1.29e-01} & \textit{1.59e-01} & \textit{1.95e-01} & \textit{2.40e-01} & \textit{2.96e-01} & \textit{3.64e-01} & \textit{4.47e-01} & \textit{5.50e-01}  \rule{0pt}{1ex} \\
\rule{0pt}{4ex}\textit{1.55e+00} & 7.10e-04 & 7.11e-04 & 6.26e-04 & 1.88e-03 & 1.29e-03 & 2.14e-03 & 2.47e-02 & 2.05e-03 & 1.06e-02 & 2.23e-03 & 1.96e-03 & 1.05e-03 & 8.26e-04 & 6.70e-04 \\
\textit{1.26e+00} & 4.81e-04 & 9.84e-04 & 4.52e-04 & 1.39e-03 & 2.64e-04 & 2.44e-03 & 1.07e-03 & 9.37e-04 & 8.60e-04 & 6.68e-04 & 7.47e-04 & 5.30e-04 & 3.89e-04 & 2.71e-04  \\
\textit{1.02e+00} & 3.32e-04 & 1.09e-03 & 2.70e-04 & 1.08e-03 & 5.16e-04 & 1.07e-03 & 1.05e-03 & 1.07e-03 & 8.69e-04 & 6.53e-04 & 5.09e-04 & 3.34e-04 & 3.49e-04 & 2.56e-04  \\
\textit{8.33e-01} & 2.23e-04 & 5.09e-04 & 1.30e-04 & 1.09e-03 & 1.95e-04 & 6.84e-04 & 6.70e-04 & 5.61e-04 & 4.93e-04 & 3.77e-04 & 3.75e-04 & 2.80e-04 & 2.83e-04 & 2.28e-04  \\
\textit{6.77e-01} & 2.41e-04 & 5.13e-04 & 1.49e-04 & 4.07e-04 & 1.39e-04 & 5.31e-04 & 5.62e-04 & 8.55e-04 & 5.11e-04 & 4.49e-04 & 3.92e-04 & 2.79e-04 & 2.40e-04 & 2.20e-04  \\
\textit{5.50e-01} & 3.24e-04 & 7.12e-04 & 3.46e-04 & 3.41e-03 & 3.90e-04 & 8.22e-04 & 1.09e-03 & 9.12e-04 & 7.50e-04 & 6.81e-04 & 4.70e-04 & 3.58e-04 & 3.91e-04 & 3.04e-04  \\
\textit{4.47e-01} & 9.20e-05 & 2.96e-04 & 2.00e-04 & 1.29e-03 & 2.92e-04 & 9.25e-04 & 5.67e-04 & 8.21e-04 & 4.89e-04 & 3.76e-04 & 3.13e-04 & 2.44e-04 & 3.32e-04 & 2.61e-04 \\
\textit{3.64e-01} & 1.03e-04 & 2.11e-04 & 5.95e-05 & 2.16e-04 & 8.84e-05 & 6.03e-04 & 2.68e-04 & 2.63e-04 & 3.38e-04 & 2.13e-04 & 1.83e-04 & 1.81e-04 & 1.76e-04 & 1.54e-04 \\
\textit{2.96e-01} & 1.16e-04 & 1.73e-04 & 6.24e-05 & 1.44e-04 & 7.23e-05 & 2.66e-04 & 2.19e-04 & 2.32e-04 & 2.35e-04 & 1.84e-04 & 1.49e-04 & 1.66e-04 & 1.74e-04 & 1.69e-04 \\
\textit{2.40e-01} & 8.08e-05 & 1.36e-04 & 7.29e-05 & 1.30e-04 & 5.64e-05 & 3.71e-04 & 1.72e-04 & 1.63e-04 & 2.06e-04 & 1.38e-04 & 1.16e-04 & 1.39e-04 & 1.63e-04 & 1.65e-04 \\
\textit{1.95e-01} & 4.92e-05 & 9.91e-05 & 3.13e-05 & 1.77e-04 & 4.19e-05 & 8.97e-05 & 2.08e-04 & 2.67e-04 & 2.45e-04 & 1.45e-04 & 1.30e-04 & 1.50e-04 & 1.76e-04 & 1.85e-04 \\
\textit{1.59e-01} & 3.82e-05 & 5.31e-05 & 2.42e-05 & 7.26e-05 & 2.94e-05 & 4.18e-04 & 1.30e-04 & 1.49e-04 & 1.56e-04 & 1.27e-04 & 1.47e-04 & 1.66e-04 & 1.93e-04 & 2.16e-04 \\
\textit{1.29e-01} & 2.65e-05 & 2.91e-05 & 2.22e-05 & 3.98e-05 & 2.16e-05 & 8.50e-05 & 8.63e-05 & 9.34e-05 & 1.06e-04 & 1.40e-04 & 1.66e-04 & 1.56e-04 & 1.90e-04 & 2.67e-04 \\
\textit{8.53e-02} & 1.67e-05 & 1.70e-05 & 1.12e-05 & 3.72e-05 & 1.13e-05 & 3.76e-05 & 1.29e-04 & 9.07e-05 & 2.23e-04 & 2.64e-04 & 2.31e-04 & 1.95e-04 & 2.54e-04 & 4.18e-04 \\
\textit{4.58e-02} & 6.07e-06 & 1.05e-05 & 5.18e-06 & 1.26e-05 & 1.09e-05 & 3.70e-05 & 1.06e-03 & 1.23e-04 & 1.89e-04 & 5.63e-03 & 7.43e-04 & 4.78e-04 & 6.71e-04 & 8.29e-04 \\ \hline
\end{tabular}}
\end{adjustbox}
\newpage
\begin{adjustbox}{angle=90}
\centering
\resizebox{0.9\paperwidth}{0.075\paperheight}{
\setlength\tabcolsep{4pt}
\begin{tabular}{|l|cccccccccccccc|}\hline
\multicolumn{1}{c}{$_{k_\parallel} \backslash ^{k_\perp}$} & \textit{6.77e-01} & \textit{8.33e-01} & \textit{1.02e+00} & \textit{1.26e+00} & \textit{1.55e+00} & \textit{1.91e+00} & \textit{2.35e+00} & \textit{2.89e+00} & \textit{3.55e+00} & \textit{4.37e+00} & \textit{5.37e+00} & \textit{6.61e+00} & \textit{8.13e+00} & \textit{1.00e+01} \\
\rule{0pt}{4ex}\textit{1.55e+00} & 5.02e-04 & 5.49e-04 & 4.66e-04 & 6.13e-04 & 7.10e-04&7.49e-04 & 1.34e-03 & 1.86e-03 & 3.09e-03 & 4.67e-03 & 7.70e-03 & 8.91e-03 & 1.39e-02 & 1.16e-02 \\
\textit{1.26e+00} & 2.54e-04 & 2.53e-04 & 2.57e-04 & 2.46e-04 & 3.66e-04 & 3.80e-04 & 5.29e-04 & 8.89e-04 & 1.79e-03 & 3.19e-03 & 6.07e-03 & 2.13e-02 & 1.74e-02 & 2.15e-02 \\
\textit{1.02e+00} & 2.43e-04 & 2.29e-04 & 2.36e-04 & 2.48e-04 & 2.66e-04 & 4.06e-04 & 6.14e-04 & 9.35e-04 & 3.31e-03 & 6.10e-03 & 1.04e-02 & 1.09e-02 & 1.91e-02 & 6.59e-02 \\
\textit{8.33e-01} & 1.87e-04 & 1.95e-04 & 1.94e-04 & 2.11e-04 & 2.91e-04 & 4.03e-04 & 7.91e-04 & 1.21e-03 & 2.23e-03 & 5.59e-03 & 2.37e-02 & 7.82e-03 & 1.73e-02 & 5.11e-02 \\
\textit{6.77e-01} & 2.08e-04 & 2.11e-04 & 2.23e-04 & 2.34e-04 & 3.46e-04 & 4.78e-04 & 7.96e-04 & 1.17e-03 & 3.35e-03 & 5.67e-03 & 8.62e-03 & 5.23e-02 & 3.28e-02 & 3.66e-02 \\
\textit{5.50e-01} & 2.93e-04 & 2.97e-04 & 2.80e-04 & 3.97e-04 & 5.24e-04 & 8.20e-04 & 1.46e-03 & 2.13e-03 & 4.36e-03 & 7.02e-03 & 1.76e-02 & 1.87e-02 & 1.58e-01 & 6.49e-02 \\
\textit{4.47e-01} & 2.42e-04 & 2.92e-04 & 2.93e-04 & 3.50e-04 & 5.29e-04 & 7.68e-04 & 1.32e-03 & 2.05e-03 & 3.78e-03 & 1.03e-02 & 1.66e-02 & 3.62e-02 & 2.91e-02 & 2.53e-01 \\
\textit{3.64e-01} & 1.68e-04 & 2.21e-04 & 2.60e-04 & 3.93e-04 & 5.61e-04 & 7.63e-04 & 1.25e-03 & 2.06e-03 & 3.70e-03 & 1.36e-02 & 1.74e-02 & 2.65e-02 & 3.25e-02 & 4.44e-02 \\
\textit{2.96e-01} & 1.97e-04 & 2.20e-04 & 3.39e-04 & 4.81e-04 & 5.99e-04 & 8.34e-04 & 1.31e-03 & 2.31e-03 & 5.94e-03 & 1.11e-02 & 2.01e-02 & 6.60e-02 & 6.60e-02 & 7.40e-02 \\
\textit{2.40e-01} & 1.78e-04 & 2.47e-04 & 3.24e-04 & 4.48e-04 & 5.98e-04 & 8.72e-04 & 1.33e-03 & 2.59e-03 & 5.83e-03 & 1.17e-02 & 2.71e-02 & 2.32e-01 & 6.35e-02 & 1.11e-01 \\
\textit{1.95e-01} & 2.17e-04 & 3.13e-04 & 3.53e-04 & 4.96e-04 & 6.98e-04 & 1.06e-03 & 1.53e-03 & 3.02e-03 & 6.04e-03 & 1.09e-02 & 2.40e-02 & 7.58e-02 & 2.46e-01 & 3.59e-02 \\
\textit{1.59e-01} & 2.46e-04 & 3.27e-04 & 4.08e-04 & 5.20e-04 & 8.08e-04 & 1.28e-03 & 1.75e-03 & 3.06e-03 & 7.07e-03 & 1.19e-02 & 2.34e-02 & 3.14e-02 & 7.79e-02 & 4.52e-01 \\
\textit{1.29e-01} & 2.99e-04 & 3.47e-04 & 4.70e-04 & 5.42e-04 & 9.46e-04 & 1.39e-03 & 2.09e-03 & 3.38e-03 & 6.97e-03 & 1.39e-02 & 2.38e-02 & 6.39e-02 & 4.89e-02 & 1.05e-01 \\
\textit{8.53e-02} & 4.42e-04 & 5.22e-04 & 6.48e-04 & 8.54e-04 & 1.34e-03 & 2.06e-03 & 3.30e-03 & 6.02e-03 & 1.07e-02 & 2.05e-02 & 3.44e-02 & 8.34e-02 & 1.04e-01 & 1.11e-01 \\
\textit{4.58e-02} & 8.50e-04 & 1.06e-03 & 1.30e-03 & 1.62e-03 & 2.58e-03 & 3.68e-03 & 5.70e-03 & 1.34e-02 & 2.16e-02 & 3.99e-02 & 5.74e-02 & 1.38e-01 & 1.58e-01 & 4.07e-01 \\
\hline
\end{tabular}}

\end{adjustbox}

\bsp	
\label{lastpage}
\end{document}